\documentclass[iop]{emulateapj}

\shorttitle{Possible Spin-Orbit Misalignment Along LOS}
\shortauthors{Kevin Schlaufman}
\submitted{Received 2009 December 24}

\begin{document}

\title{Evidence of Possible Spin-Orbit Misalignment Along the Line of Sight in
Transiting Exoplanet Systems}

\author{Kevin C. Schlaufman\altaffilmark{1}}
\affil{Astronomy and Astrophysics Department, University of
California, Santa Cruz, CA 95064}
\email{kcs@ucolick.org}
\altaffiltext{1}{NSF Graduate Research Fellow}

\begin{abstract}

There has lately been intense interest in the degree of alignment
between the orbits of transiting exoplanets and the spin of their hosts
stars.  Indeed, of 26 transiting exoplanet systems with measurements
of the Rossiter-McLaughlin (RM) effect, eight have now been found
to be significantly spin-orbit misaligned in the plane of the sky.
Unfortunately, the RM effect only measures the angle between the orbit
of a transiting exoplanet and the spin of its host star projected in the
plane of sky, leaving unconstrained the compliment misalignment angle
between the orbit of the planet and the spin of its host star along the
line of sight.  I use a simple model of stellar rotation benchmarked
with observational data to statistically identify ten exoplanet systems
from a sample of 75 for which there is likely a significant degree of
misalignment along the line of sight between the orbit of the planet
and the spin of its host star.  I find that HAT-P-7, HAT-P-14, HAT-P-16,
HD 17156, Kepler-5, Kepler-7, TrES-4, WASP-1, WASP-12, and WASP-14 are
likely spin-orbit misaligned along the line of sight.  All ten systems
have host stellar masses $M_{\ast}$ in the range $1.2~M_{\odot} \lesssim
M_{\ast} \lesssim 1.5~M_{\odot}$, and the probability of this occurrence
by chance is less than one in ten thousand.  In addition, the planets in
the candidate misaligned systems are preferentially massive and eccentric.
I combine the distribution of misalignment in the plane of the sky from RM
measurements with the distribution of misalignment along the line of sight
from this analysis, and the result suggests that transiting exoplanets
are more likely to be aligned along the line of sight than expected
given predictions for the spin-orbit misalignment distribution of a
population produced entirely by planet-planet scattering or Kozai cycles
and tidal friction.  For that reason, there are likely two populations
of close-in exoplanet systems: a population of aligned systems and a
population of apparently misaligned systems in which the processes that
lead to misalignment or to the survival of misaligned systems operate
more efficiently in systems with massive stars and planets.

\end{abstract}


\section{Introduction}

The degree of alignment $\psi$ between the orbit of a transiting
exoplanet and the spin of its host star can provide insight into
the system's formation.  In the Solar System, the angle $\psi$
between the planets' orbits and the spin of the Sun is small, about
7$^{\circ}$ \citep[e.g.][]{bec05}.  That fact provided a key piece
of evidence supporting the earliest models of planet formation in
the same disk in which the Sun formed.  Simple models of close-in
exoplanet formation involving assembly near the ice line and quiescent
migration through a gaseous protoplanetary disk predict that orbits
should have only negligible misalignments with the spin of the host star
\citep[e.g.][]{lin96,ida08a,ida08b}.  Alternatively, simple models without
migration but involving dynamical scattering between newly formed planets
and planetesimals at several AU predict that non-negligible misalignments
$\psi$ can occur \citep[e.g.][]{for01,yu01,pap01,ter02,mar02,cha08,jur08}.
In reality, there will be some degree of migration in all massive
planetary systems \citep{war97}, and in certain situations migration
in multiple systems can still lead to non-negligible misalignments
\citep{tho03}.  Moreover, any multiple systems that migrate will
likely be dynamically stable before the dissipation of their gaseous
parent protoplanetary disk, but dynamically unstable after dissipation
\citep{ter07,ogi09}.  The Kozai effect \citep{koz62} and tidal processes
might also contribute \citep[e.g.][]{win05,fab07,nag08,pon09a}.
Despite these complications, the frequency of misaligned systems will
provide another constraint on models of planet formation.

\begin{figure}
\plotone{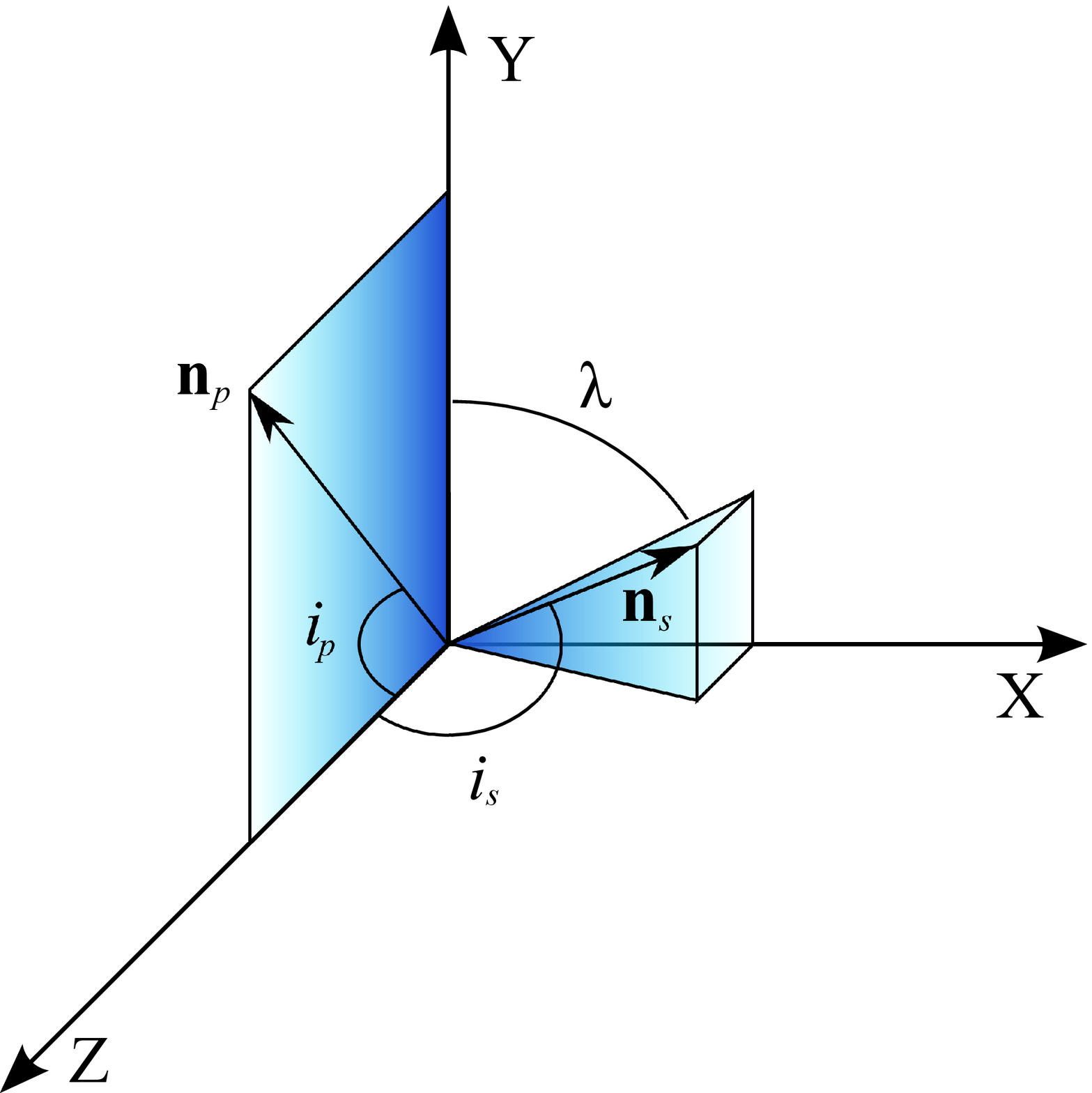}
\caption{The geometry of transiting exoplanet systems following the
conventions described in \citet{fab09}, with the $z$-axis oriented toward
the observer.  The exoplanet orbit is defined by the vector $\mathbf{n}_p$
and is inclined at an angle $i_p$ to the observer's line of sight.  The
host stellar spin is defined by the vector $\mathbf{n}_s$ and is inclined
at an angle $i_s$ to the observer's line of sight.  Transit measurements
can determine $i_p$, and measurements of the Rossiter-McLaughlin effect
can reveal $\lambda$, the angle in the $xy$-plane between the projection
of $\mathbf{n}_p$ and $\mathbf{n}_s$ \citep{ros24,mcl24,oht05}.  The angle
$i_s$ is difficult to measure for individual stars, but as I show in this
paper, it may be possible to identify systems in which $i_s$ and $i_p$
are misaligned by a large degree.\label{fig01}}
\end{figure}

I describe the geometry of transiting exoplanet systems according to
the conventions of \citet{fab09} in Figure~\ref{fig01}.  An exoplanet's
orbit is defined by the vector $\mathbf{n}_p$ and is inclined relative
to the line of sight at an angle $i_p$ that can be measured through
transit photometry.  Note that the fact that an exoplanet is observed
to transit requires that $i_p \approx 90^{\circ}$.  The host star's
spin is defined by the vector $\mathbf{n}_s$ and is inclined at an
angle $i_s$ relative to the line of sight.  The angle $\lambda$ is
the angular distance between the projections of $\mathbf{n}_p$ and
$\mathbf{n}_s$ in the plane of the sky, and is observable through the
Rossiter-McLaughlin (RM) effect \citep{ros24,mcl24,oht05}.  The RM effect
-- the detection of an anomalous blueshift and redshift superimposed
on the stellar reflex in radial velocity measurements during transit
-- has lately been used to measure $\lambda$ for several transiting
exoplanet systems \citep[e.g.][and references therein]{fab09}.  Indeed,
the latest RM measurements described in Triaud et al. (2010, submitted)
show that eight of the 26 systems with RM measurements are significantly
spin-orbit misaligned in the plane of the sky.  Though $i_s$ is not
constrained by the RM effect, if all of $i_p$, $i_s$, and $\lambda$
are known, the calculation of the true deprojected angle $\psi \equiv
\arccos(\mathbf{n}_p \cdot \mathbf{n}_s)$ between $\mathbf{n}_p$ and
$\mathbf{n}_s$ is trivial.

The inclination $i_s$ to the line of sight of stellar spin is difficult
to measure for individual stars.  If the rotation period of a star
is known through photometric observations of starspots moving across
its surface, and if the projected rotational velocity $v\sin{i}$ has
been measured, then $i_s$ is easy to compute \citep[e.g.][]{win07}.
Astrosiesmology measurements of $i_s$ are also possible in special
circumstances \citep{giz03}.  Unfortunately, period measurements are
resource-intensive and require well-sampled and precise photometric
time series.  They may also be biased toward stars with many starspots
and therefore biased toward more active and younger stars.  On the other
hand, the projection of stellar rotation velocity along the line of
sight $v\sin{i}$ can be measured cheaply by examining the rotational
broadening of absorption lines in a single high-resolution stellar
spectrum.  If the radius of the star is known from other means, it is
easy to forward-model the observable quantity $v\sin{i}$ from the stellar
rotation period $P_{\ast}$ predicted from theoretical or empirical models
given a distribution for $i$.  Monte Carlo simulations can then derive the
distribution of possible $v\sin{i}$ measurements.  That distribution can
then be compared to observational data for transiting exoplanet systems
to determine if the observations are consistent with the alignment of
a transiting planet's orbit with the spin of its host star.

Calculating the rotation period of Sun-like stars as a function of mass
and time is a tractable problem for stars older than about 650 Myr.
First, the rotation period of isolated Sun-like stars monotonically slows
down as they age \citep{kra67}.  Second, even though a population of
Sun-like stars is formed with a broad distribution of periods $P_{\ast}$
\citep{att92,cho96}, by the time that population is the age of the Hyades
and Praesepe (about 650 Myr) the broad distribution has converged to
a well-defined function of mass \citep{tas00,irw09}.  Two classes of
physical models have been proposed to explain this early-time behavior:
magnetic breaking by protostellar disks \citep[e.g.][]{koe91,shu94,col95}
and accretion-driven stellar winds \citep[e.g.][]{mat05}.  Though the
physical process that produces this effect in the first $\sim 650$ Myr
is not yet known, the data \citep{rad87,pro95,sch07} and theoretical
models \citep{kep95,tas00} are in agreement on the rotational
evolution of Sun-like stars after this time.  Indeed, Sun-like
stars likely lose angular momentum through magnetized stellar winds
\citep{web67,mes68,kaw88}, and their rotation slows as $P_{\ast} \propto
t^{1/2}$, as described observationally as $v\sin{i} \propto t^{-1/2}$
by \citet{sku72}.

In this paper, I use a simple empirical model to calculate the rotation
period of Sun-like stars as a function of mass and age.  I combine that
model with a Monte Carlo simulation to marginalize over the uncertain
masses, radii, and ages of a control sample of 866 stars from the
Spectroscopic Properties of Cool Stars (SPOCS) catalog of \citet{val05}
to determine the expected range of projected rotation $v\sin{i}$ for each
star in the control sample under the assumption that the inclination
distribution of the SPOCS sample is isotropic.  Likewise, I perform a
similar calculation for the host stars of transiting exoplanets, this
time assuming alignment between the host star's spin and the orbit of its
exoplanet.  I identify those transiting exoplanet systems in which the
disagreement between the $v\sin{i}$ predicted from the simple empirical
model under the assumption of spin-orbit alignment and the observed
$v\sin{i}$ is larger than the equivalent disagreement for any of the
866 stars in the SPOCS control sample.  In particular, those systems
that have anomalously small $v\sin{i}$ measurements relative to the
measurement errors and the width of simulated $v\sin{i}$ distribution are
possibly spin-orbit misaligned along the line of sight.  I describe the
details of my calculation in \S2.  I discuss my results in the context
of close-in planet formation in \S3.  I summarize my findings in \S4.

\section{Analysis}

I combine the well defined mass-rotation period relation in the Hyades
and Praesepe \citep[e.g.][]{irw09} with the relation $P_{\ast} \propto
t^{1/2}$ \citep[e.g.][]{web67,mes68,kaw88} to compute the expected
rotation periods of Sun-like stars as a function of mass and age.  I use a
Monte Carlo simulation to transform that function into a $v\sin{i}_{sim}$
distribution by marginalizing over the uncertain distributions of
stellar mass $M_{\ast}$, radius $R_{\ast}$, age $\tau_{\ast}$, and
the applicable inclination distribution for the sample.  As a result,
I can compare the observed $v\sin{i}_{obs}$ measurements with the mean
$\overline{v\sin{i}}_{sim}$ estimate from the Monte Carlo simulation
to determine the degree of agreement between the two.  In order to
determine the degree of disagreement that can be attributed to random
effects, I compare the measured values of $v\sin{i}_{obs}$ of Sun-like
stars in the Solar Neighborhood from the SPOCS catalog with the mean
$\overline{v\sin{i}}_{sim}$ from the Monte Carlo simulation given
the stellar parameters of those stars and the standard isotropic
inclination distribution.  In this way, I use the control sample of
SPOCS stars to determine a threshold level of disagreement that can be
expected between the simple empirical model and observation given the
observational uncertainties and the imperfections of the model.  I then
do a similar calculation for the host stars of transiting exoplanets,
this time assuming alignment between the spin of the host star and
the orbit of its exoplanet.  I identify those exoplanet host stars
in which the predicted $\overline{v\sin{i}}_{sim}$ disagrees with the
observed value $v\sin{i}_{obs}$ relative to the measurement errors by
an amount greater then an equivalent disagreement for any of the 866
stars in the control sample.  Systems with $v\sin{i}_{obs}$ slower than
$\overline{v\sin{i}}_{sim}$ by an amount large relative to the quoted
errors in $v\sin{i}_{obs}$ and the width of the simulated $v\sin{i}_{sim}$
distribution given the uncertainty in stellar mass, radius, and age are
either systems with rotation properties unlike any star in the control
sample or in which the inclination of the transiting planet's orbit
$i_p$ and the inclination of the host star's spin $i_s$ are misaligned.
Since the SPOCS sample is representative of the sample of transiting
exoplanet host stars, the former is very unlikely.  For that reason,
the best explanation of the anomalously slow projected rotation is
spin-orbit misalignment along the line of sight.

\subsection{SPOCS Control Sample}

I examine a control sample of stars from the SPOCS catalog of
\citet{val05} to quantify the scatter about the simple model.  The SPOCS
sample selection is described in detail in \citet{mar04}.  In summary,
the bulk of the sample was initially selected from the Hipparcos catalog
such that each star has $B-V>0.55$, is no more than three magnitudes
above the main sequence, is not a known spectroscopic binary, and has
no stellar companion within two arcseconds.  Subsequent spectroscopic
measurements of CaII H and K were then used to downselect to those stars
with ages $\tau_{\ast} \gtrsim 2$ Gyr, though an additional 100 stars
with ages between 50 and 500 Myr were later added in.  As a result, the
SPOCS sample has both young and evolved stars as indicated in Figure 14
of \citet{val05}.

In addition, previously unknown spectroscopic binaries were removed from
the SPOCS candidate list, so there are no spectroscopic binaries in the
final catalog.  Short-period low-mass binary companions or exoplanets
would have been readily apparent in the original California-Carnegie
planet search and would have been announced as such long ago.  For those
reasons, the rotation properties of stars in SPOCS sample should be
unaffected by close companions.  Moreover, any overlap between the
SPOCS sample and the hosts stars of transiting exoplanets is negligible.
Collectively, all of these selections and observations suggest that the
SPOCS sample is a fair control sample for this analysis.

\subsection{Detailed Description of the Monte Carlo}

\begin{figure}
\plotone{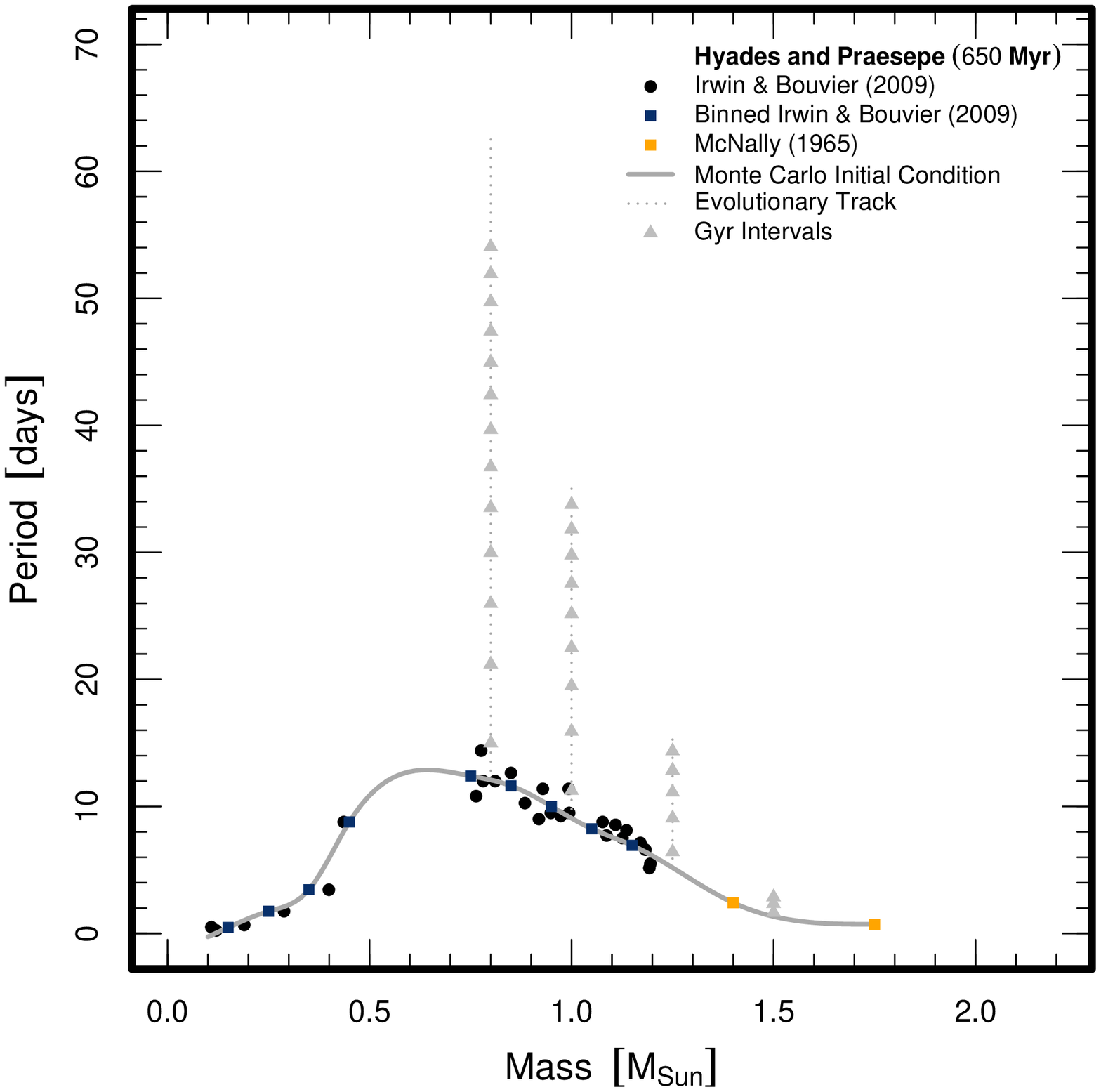}
\caption{Stellar rotation period of Sun-like stars as a function of mass
and age.  Though a population of Sun-like stars is formed with a wide
range of rotation periods $P_{\ast}$ \citep{att92,cho96}, by the time
that population is the age of the Hyades and Praesepe period is only a
function of mass \citep[e.g.][]{tas00,irw09}.  Main sequence stars lose
angular momentum and evolve from from this initial condition, likely
through a magnetized stellar wind \citep[e.g.][]{web67,mes68,kaw88}.
As a result, stellar periods increase on the main sequence as they age
\citep[e.g.][]{kra67} as $P_{\ast} \propto t^{1/2}$, in accordance with
the empirical relation $v\sin{i} \propto t^{-1/2}$ noted by \citet{sku72}.
I binned the available Hyades and Praesepe data \citep{rad87,pro95,sch07}
as summarized in \citet{irw09} (black points) in bins of 0.1 $M_{\odot}$
(blue squares), supplemented it with average rotation values for higher
mass field stars \citep{mcn65} (orange squares), and used natural cubic
spline interpolation to compute a relationship between mass and period for
stars between $0.2~M_{\odot} \lesssim M_{\ast} \lesssim 1.75~ M_{\odot}$
(solid gray curve).  From this initial condition, I evolve the rotation
period as $P_{\ast} \propto t^{1/2}$ (dotted gray vertical lines).
The length of the vertical lines correspond to the main sequence lifetime
of a star of the given mass, and I denote Gyr intervals (from 1 Gyr to
13 Gyr) with gray triangles along each vertical line.\label{fig02}}
\end{figure}

I model the initial mass-period relation of a 650 Myr population of
Sun-like stars by binning the available Hyades and Praesepe data
\citep{rad87,pro95,sch07} as summarized in \citet{irw09} in 0.1
$M_{\odot}$ bins.  The \citet{irw09} data only extends up to about
1.2 $M_{\odot}$, so I supplement it with average rotation periods for
more massive field stars as presented in \citet{mcn65}.  I use natural
cubic splines to interpolate the binned and supplemented data and find
that it minimizes the sum of square residuals relative to high-order
polynomial interpolation.  I determine the expected rotation period
of a Sun-like star as a function of mass and age by evolving the
initial condition set by the stellar mass according to the relation
\citep[e.g.][]{web67,mes68,kaw88}

\begin{eqnarray}\label{eq1}
P_{\ast}(M_{\ast},\tau_{\ast}) & = & P_{\ast,0}(M_{\ast}) \left(\frac{\tau_{\ast}}{\mbox{650 Myr}}\right)^{1/2}
\end{eqnarray}

\noindent
where $P_{\ast,0}(M_{\ast})$ can be approximated as a fifth-order
polynomial in $M_{\ast}$ with coefficients in increasing order
(5.558509,-77.843724,357.387761,-502.288156,282.908686,-56.329952).
I summarize my technique in Figure~\ref{fig02}.

The model described above is likely too simplistic, the relevant
input parameters (stellar mass, radius, and age) are all uncertain,
and the output (stellar period) is difficult to observe.  For those
reasons, I use a Monte Carlo simulation to determine the range of
observable $v\sin{i}_{sim}$ expected for a sample of stars given
mass, radius, and age as well as the uncertainties in each of those
quantities plus the appropriate inclination distribution for that
sample.  In order to determine the degree of disagreement between
the $\overline{v\sin{i}}_{sim}$ predicted from the simple empirical
model and the observed $v\sin{i}_{obs}$ that can be expected given the
observational uncertainties and the imperfections of the simple empirical
model, I use the control sample of stars from the SPOCS catalog.

For each star in the SPOCS catalog, I use the estimated stellar mass and
uncertainty $M_{\ast}$ and $\sigma_{M}$ (column 9 of Table 9), estimated
stellar radius and uncertainty $R_{\ast}$ and $\sigma_{R}$ (column 8
of Table 9), and age range $\Delta\tau$ (column 14 of Table 9).  In the
Monte Carlo, I sample each star's mass from a uniform distribution in mass
between $M_{\ast}-\sigma_{M}/2$ and $M_{\ast}+\sigma_{M}/2$, its radius
from a uniform distribution in radius between $R_{\ast}-\sigma_{R}/2$
and $R_{\ast}+\sigma_{R}/2$, and its age from a uniform distribution
in age between the given lower and upper age limits.  I sample the
inclination to the line of sight of each star's rotation axis from
the standard random distribution $i \sim \arccos(1-U)$, where $U$
is drawn uniformly from the interval [0,1].  Using these parameters,
I evolve the initial period given by the randomly selected mass and the
initial period-mass relation described in Figure~\ref{fig02} to the
randomly selected age of the system according to Equation~\ref{eq1}.
I then compute $v\sin{i}_{sim}$ using the randomly generated radius
of the star and its randomly selected inclination.  I repeat this
process 1000 times and thereby derive for each star the distribution of
possible $v\sin{i}_{sim}$ values, its mean $\overline{v\sin{i}}_{sim}$,
as well as the width of the distribution $\sigma_{sim}$.  I compare the
predicted $\overline{v\sin{i}}_{sim}$ with the observed $v\sin{i}_{obs}$
relative to the width of the Monte Carlo distribution $\sigma_{sim}$
and the error in the observed $v\sin{i}_{obs}$ measurement $\sigma_{obs}$
through the rotation statistic $\Theta$:

\begin{eqnarray}\label{eq2}
\Theta & \equiv & \frac{\overline{v\sin{i}}_{sim}-v\sin{i}_{obs}}{\sqrt{\sigma_{obs}^2+\sigma_{sim}^2}}
\end{eqnarray}

\noindent
Note that large positive (negative) values of $\Theta$ indicate slower
(faster) than expected $v\sin{i}$ values.  Stars that are well fit by
the model will have small absolute values of $\Theta$.

I repeat this calculation on each of the 866 stars from the SPOCS
catalog with $M_{\ast} < 1.5~M_{\odot}$, assuming the standard isotropic
distribution of inclination.  This distribution of $\Theta$ is the null
hypothesis that I will use to compare to the distribution of $\Theta$
measured in the transiting exoplanet systems under the assumption of
spin-orbit alignment.

I collected the properties of the 80 transiting exoplanets systems
known as of May 2010, of which the 75 listed in Table~\ref{tbl-1}
have measured mass, radius, $v\sin{i}_{obs}$, and orbital inclination.
In this case, the stellar masses and radii are more precisely measured
than those available in the SPOCS catalog, and the inclination of each
exoplanet's orbit is known from transit photometry.  I again use a Monte
Carlo simulation to determine the range of observable $v\sin{i}_{sim}$.
I use the measured values for each star's mass and radius, and I sample
each star's age uniformly between lower and upper bounds for its age
from the literature.  If there are no age estimates available in the
literature, I sample the age of the star uniformly between 650 Myr (the
age of the Hyades and Praesepe) and its main sequence lifetime.  If there
are only upper limits on $v\sin{i}_{obs}$ in the literature, I assume
that the true value corresponds to the published upper-limit, as this is
the most conservative way to approach a search for stars with anomalously
slow projected rotation.  In those situations, I assume that the error
in $v\sin{i}_{obs}$ is 1 km s$^{-1}$.  I assume that the inclination of
the host star's spin relative to the line of sight $i_s$ is the same the
inclination of the exoplanet's orbit $i_p$.  I repeat this process 1000
times and compute the rotation statistic $\Theta$ for each stellar host;
I report the results of this calculation in Table~\ref{tbl-1}.

\subsection{Results}

\begin{figure}
\plotone{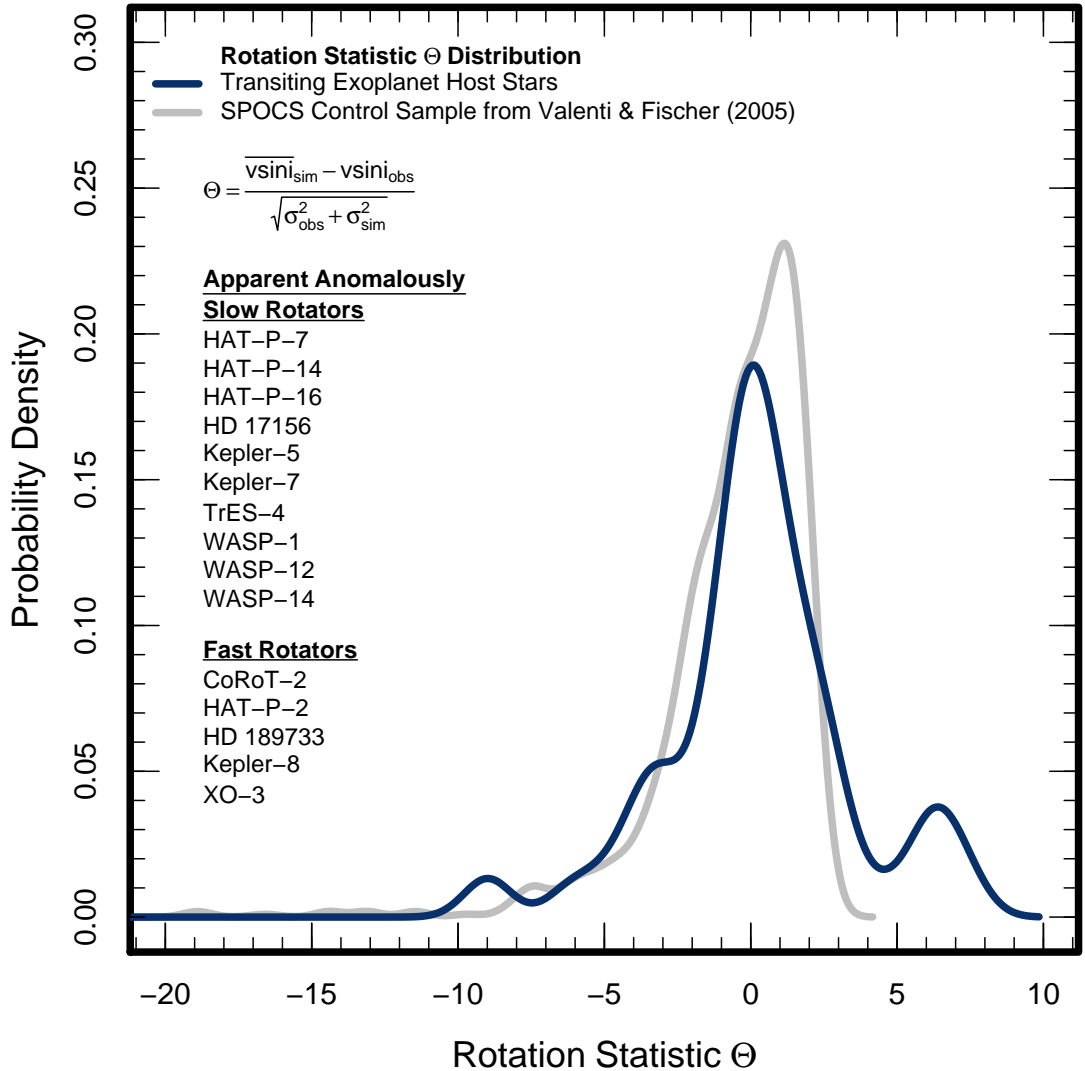}
\caption{The distribution of the rotation statistic $\Theta$ for
stars that host transiting exoplanets (solid blue line) compared
to a control sample from the SPOCS catalog of \citet{val05} (solid
gray line).  I define $\Theta \equiv \left(\overline{v\sin{i}}_{sim} -
v\sin{i}_{obs}\right)/\sqrt{\sigma_{obs}^2+\sigma_{sim}^2}$ such that
stars that have a $v\sin{i}$ smaller (larger) than expected given their
mass, age, radius, and the uncertainties in those parameters have positive
(negative) values of $\Theta$.  By comparison with the control sample,
transiting exoplanet host stars that have $\Theta > 2.9$ have $v\sin{i}$
so small that they cannot be explained by evolving the Hyades and Praesepe
data according to the $P_{\ast} \propto t^{1/2}$ relation given the
system parameters and uncertainties and assuming alignment between the
exoplanet orbit and the spin of its host.  I argue that unless these
stellar hosts of transiting exoplanets are slower rotators relative
to the model than any of the 866 stars in the SPOCS control sample,
the most natural explanation for this result is that $i_s$ and $i_p$
are misaligned.\label{fig03}}
\end{figure}

I plot the distribution of $\Theta$ for both samples in
Figure~\ref{fig03}.  The $\Theta$ distribution of the control sample peaks
near $\Theta = 0$, indicating that the simple empirical model is accurate
in that the expected residual between the model and the observations is
close to zero.  The distribution of $\Theta$ falls off quickly as $\Theta$
increases, and there are no control stars from the SPOCS catalog with
$\Theta > 2.9$.  However, there does exist a tail of stars with much
larger than expected $v\sin{i}$ values, and therefore large negative
$\Theta$ values.  In other words, some stars in the control sample of
SPOCS stars are fast rotators; this is to be expected in any sample of
Sun-like field stars.  On the other hand, there is no tail of SPOCS stars
rotating arbitrarily more slowly than the simple empirical model predicts.
For that reason, host stars of exoplanets that are observed to have
$\Theta > 2.9$ are apparently unlike any of the 866 stars in the control
sample -- either they rotate more slowly relative to the model than any
star in the control sample or their anomalously slow projected rotation is
due to a pole-on view of the star and therefore spin-orbit misalignment.

\begin{figure}
\plotone{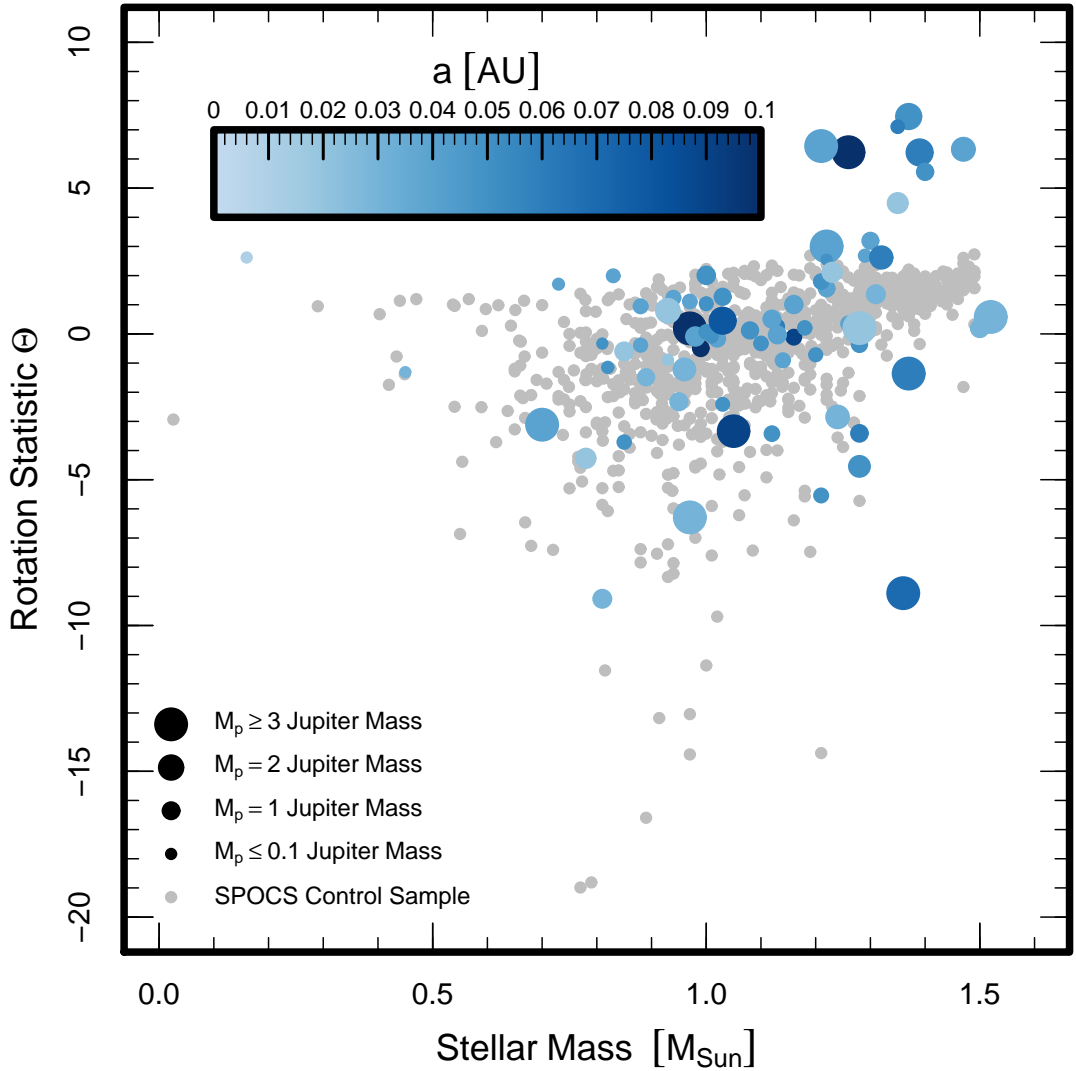}
\caption{The rotation statistic $\Theta$ as a function of stellar
mass for both stars that host transiting exoplanets (blue points) and
control stars from the SPOCS catalog (gray points).  Note that stars with
$\Theta > 2.9$ -- anomalously slow projected rotation assuming spin-orbit
alignment -- are preferentially massive.  In the sample of 75 systems
in Table~\ref{tbl-1}, the probability that all ten misaligned systems
have host stellar mass in the range $1.2~M_{\odot} \lesssim M_{\ast}
\lesssim 1.5~M_{\odot}$ by chance is less than one in ten thousand.
In addition, the exoplanet host stars with $\Theta > 2.9$ host planets
with a range of semimajor axes, not just those at small semimajor axes
where tidal forces are most important.  In any case, the rotation period
of each exoplanet host star with $\Theta > 2.9$ is expected to be longer
than the orbital period of its planet, so those stars are unlikely to
transfer angular momentum to their planets.\label{fig04}}
\end{figure}

I define fast rotators as those stars with $\Theta < -5.2$, corresponding
to the fastest 5\% of the $\Theta$ distribution for the SPOCS control
sample.  Note that the fraction of fast rotators in each sample is
statistically indistinguishable -- about $5\% \pm 3\%$ in the SPOCS sample
and about $7\% \pm 3\%$ in the transiting exoplanet host star sample.
The consistency of the two samples in the fraction of stars that are fast
rotators is in sharp contrast to the disparity at the other end of the
distribution.  Also, the peak of the $\Theta$ distribution of transiting
exoplanet systems is offset to more negative values of $\Theta$ (and
therefore faster rotation) than the field star sample.  Nevertheless,
the distribution of the sample mean of $\Theta$ for the 90\% of the
SPOCS control sample with  $-5.2 < \Theta < 2$ is $\overline{\Theta} =
-0.25 \pm 0.19$; for the host stars of transiting exoplanets an equivalent
calculation yields $\overline{\Theta} = -0.49 \pm 0.19$.  I plot $\Theta$
as a function of stellar mass for both samples in Figure~\ref{fig04}.
There is a very sharp cutoff in the control sample at $\Theta \approx 2$
at all masses, and this sharp cutoff is due to lack of a constraint on
the inclination of individual stars in the control sample.  As a result,
$\sigma_{sim}^2$ in Equation~\ref{eq2} can be large, resulting in small
values of $\Theta$.

I find that HAT-P-7, HAT-P-14, HAT-P-16, HD 17156, Kepler-5, Kepler-7,
TrES-4, WASP-1, WASP-12, and WASP-14 have $\Theta>2.9$, indicating
$v\sin{i}_{obs}$ too small to be explained by the uncertainties in their
masses, radii, or ages if each star's spin and its exoplanet's orbit
have the same inclination to the line of sight.  Unless the host stars
of these transiting exoplanets are slower rotators relative to the model
prediction than any of the 866 stars in the control sample from the SPOCS
catalog, the most natural explanation for these discrepancies is that
the inclination of the spin of these host stars $i_s$ is significantly
different that the inclination of their exoplanet's orbit $i_p$.
If these anomalously slow projected rotation values are the result of
spin-orbit misalignment, then the difference between $v\sin{i}_{obs}$
and $\overline{v\sin{i}}_{sim}$ can be used to compute the four possible
degenerate values of $i_s$ (because $\sin{i}$ is not one-to-one for
$0^{\circ} \leq i < 360^{\circ}$).  I report these values for the
potentially spin-orbit misaligned systems in Table~\ref{tbl-2}.  I also
identify CoRoT-2, HAT-P-2, HD 189733, Kepler-8, and XO-3 as fast rotators.
Note that CoRoT-2 \citep{alo08} and HD 189733 \citep{bou05} are known
to be more active than the Sun, and therefore potential fast rotators.
HAT-P-2 has been observed to have a large radial velocity jitter, which
may indicate an active and possibly quickly rotating star \citep{bak07a}.

All ten candidate misaligned systems have host stellar mass $M_{\ast}
\gtrsim 1.2~M_{\odot}$.  Again, either these stars have rotation
properties unlike any of the 276 control stars in the mass range
$1.2~M_{\odot} \leq M_{\ast} \leq 1.5~M_{\odot}$, or these stars are
observed pole-on and therefore spin-orbit misaligned.

\subsection{Verification of the Robustness of This Result}

In order to verify the robustness of this result, I have repeated my
calculations with several different sets of input parameters.  Indeed,
there are two potentially uncertain inputs to my simple empirical model:
the power-law exponent $\alpha$ in the relation $P_{\ast} \propto
t^{\alpha}$ and the ages of the stars in the SPOCS control sample.
It is certain that the rotation of isolated Sun-like stars slows
down with time, so $\alpha > 0$ always.  To determine if my result is
sensitive to $\alpha$, I have repeated my calculation for $\alpha =
1/3$, $\alpha = 2/3$, and $\alpha = 1$.  If $\alpha = 1/3$, then all
of the systems identified with anomalously slow projected rotation in
\S2.3 are again identified as anomalously slow projected rotators.
If $\alpha = 2/3$, eight of the original ten systems are identified
as anomalously slow projected rotators (HAT-P-7, HAT-P-14, Kepler-5,
Kepler-7, HD 17156, TrES-4, WASP-12, and WASP-14).  If $\alpha = 1$,
only three of the original ten systems are identified as anomalously
slow projected rotators (HAT-P-7, HAT-P-14, and WASP-14).

I have also redone my calculation using a completely uninformative
prior on the ages of the stars in SPOCS sample.  That is, I assumed that
the age of each star was uniformly distributed between 650 Myr and its
main sequence lifetime.  Again, all ten systems identified in \S2.3 as
anomalously slow projected rotators and therefore possibly spin-orbit
misaligned are again identified in this case.

Angular momentum exchange between a host star and its close-in planetary
companion is also unlikely to significantly slow the rotation of the
host star and produce an anomalously slow projected rotation velocity.
For that to occur, the transfer of angular momentum from host star to
exoplanet via tidal forces requires the rotation period of the star to
be shorter than the orbital period of the planet.  Indeed, the median
orbital period of the transiting planets orbiting the stars identified
as anomalously slow projected rotators is $P \approx 3.2$ days, while
the median expected rotation period of their host stars is $P_{\ast}
\approx 6.3$ days.

\begin{figure}
\plotone{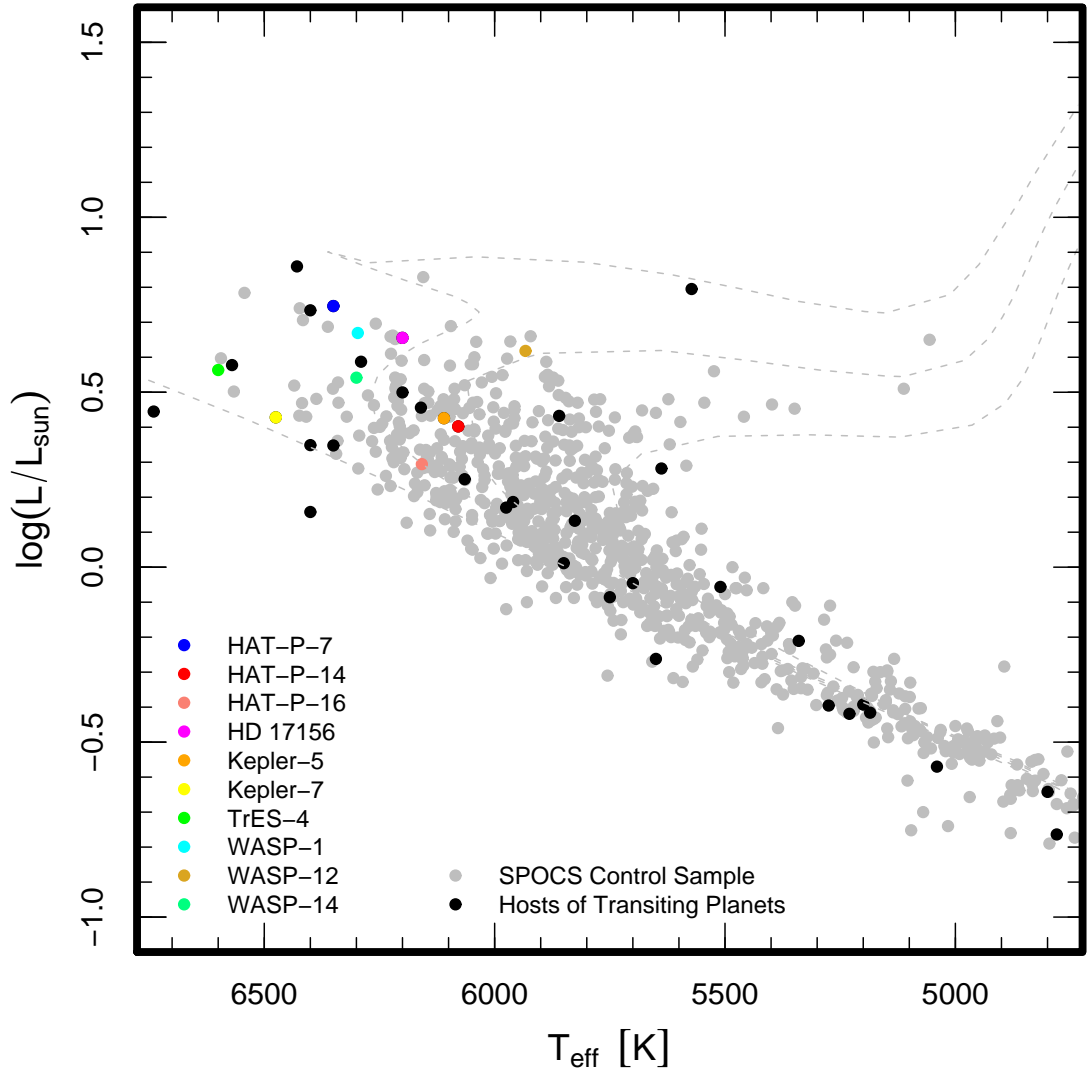}
\caption{A theoretical HR diagram of the control sample of SPOCS stars
(gray points) overplotted with the anomalously slow projected rotators
identified in this analysis.  I also plot other hosts of transiting
exoplanets with precise trigonometric parallaxes (black points).
The dashed lines are Padova isochrones for ages 0.1, 3, 5, and 10 Gyr
\citep{mar08}.  There are both young and evolved stars in the control
sample, so the SPOCS sample is a fair control sample in regards to
evolutionary state.  In addition, the fact that there are evolved hosts
of transiting exoplanets that are not identified as anomalously slow
projected rotators suggests that the rotational evolution of evolved
main sequence stars alone cannot explain the anomalously slow projected
rotation identified in some systems.\label{fig05}}
\end{figure}

The presence of slightly evolved stars among the host stars of transiting
exoplanets will also not produce false identifications as anomalously slow
projected rotators.  To see why, note that there are also slightly evolved
stars in the control sample of SPOCS stars and recall that the threshold
$\Theta$ required to identify a star as an anomalously slow projected
rotator was set by the maximum disagreement between model and observation
in the control sample.  In other words, the presence of slightly
evolved stars has already been accounted for in setting the threshold.
I plot in Figure~\ref{fig05} the distribution of the SPOCS sample in
a theoretical HR diagram along with those transiting exoplanet hosts
with trigonometric parallaxes.  I also indicate the locations of those
stellar hosts that I identified as anomalously slow projected rotators and
therefore potentially spin-orbit misaligned systems.  Figure~\ref{fig05}
shows that many of the systems I identified as anomalously slow projected
rotators appear to be evolving off of the zero-age main sequence (ZAMS).
Nevertheless, the fact that there are many stars in the control sample
of SPOCS stars in similar evolutionary states suggests that the effects
of stellar evolution on the rotation properties of evolved stars have
already been accounted for in my analysis.  Moreover, there are several
hosts of transiting exoplanets that have evolved off of the ZAMS that I
do not identify as anomalously slow projected rotators.  Collectively,
these facts suggest that effects of stellar evolution cannot explain the
rotation properties of the ten systems I identified as anomalously slow
projected rotators.

Additional selection effects in either the SPOCS control sample or the
transiting exoplanet sample are unlikely to affect this result.  Indeed,
the fact that SPOCS stars were preferentially selected to be those stars
with small enough $v\sin{i}$ to enable high precision Doppler radial
velocity measurements biases my control sample to smaller $v\sin{i}$
at fixed stellar mass and age.  However, since the candidate spin-orbit
misaligned stars appear to have anomalously slow projected rotation, the
bias acts against the false identification of a star as an anomalously
slow projected rotator.  Though magnitude-limited transit surveys are
biased toward stars slightly more massive than the Sun, there are
276 stars in the control sample with $M_{\ast} \geq 1.2~M_{\odot}$
to compare with.  In addition, though targeted transit surveys are
biased toward less active stars, the control sample of SPOCS stars in
similarly biased.  Transit surveys should also not be biased toward
evolved stars, as though both transit probability and transit duration
increase linearly with host stellar radius (favoring evolved systems),
transit depth decreases quadratically with host stellar radius (favoring
unevolved systems).  For that reason, the chance of detecting a transiting
planet is roughly insensitive to host stellar radius.

\subsection{Comparison with Rossiter-McLaughlin Measurements}

For the anomalously slow projected rotators identified in \S2.3,
I list in Table~\ref{tbl-2} the inclination to the line of sight
of the stellar rotation axis if the anomalous projected rotation is
due to spin-orbit misalignment along the line of sight.  I also note
published RM measurements for four of the systems I identify as apparent
slow projected rotators; HAT-P-7 and WASP-14 has been identified as a
spin-orbit misaligned system \citep{win09b,joh09}, while HD 17156 and
TrES-4 are consistent with spin-orbit alignment in the plane of the
sky \citep{nar09,nar10}.

Though HD 17156 is consistent with spin-orbit alignment in the plane of
the sky from RM measurements, my technique measures misalignment along the
line of sight.  There is other circumstantial evidence that HD 17156 is
misaligned along the line of sight.  \citet{fis07} measured a $v\sin{i}
= 2.6$ km s$^{-1}$, a stellar radius $R_{\ast} = 1.47~R_{\odot}$ and
used Ca II H and K to infer a rotation period $P_{\ast} = 12.8$ days.
If the system has $i = 90^{\circ}$, then a period $P_{\ast} = 12.8$
days and the inferred stellar radius $R_{\ast} = 1.47~R_{\odot}$ implies
$v\sin{i} = 5.8$ km s$^{-1}$ (as compared to $v\sin{i} = 6.9$ km s$^{-1}$
expected based on my simple model).  Clearly, it's possible that in this
case the activity indicator gives a spurious period estimate; still,
if the discrepancy is due to misalignment then $i_s = 26^{\circ}$
(as compared to $i_s = 22^{\circ}$ degrees from my measurement).

TrES-4 is consistent with alignment in the plane of the sky from
measurements of the RM effect.  However, that is a different angle than
that measured through my analysis.  It could be that the geometry of
the TrES-4 system is apparently aligned in the plane of the sky but
misaligned along the line of sight; alternatively, TrES-4 could simply
be an anomalously slow rotator.

I find that ten of the 75 systems in Table~\ref{tbl-1} have anomalously
slow projected rotation and are likely spin-orbit misaligned along the
line of sight.  In contrast, the latest Rossiter-McLaughlin measurements
reported and summarized in Triaud et al. (2010, submitted) suggest
that eight of the 26 systems with RM measurements have significant
spin-orbit misalignment in the plane of the sky.  The higher fraction
of systems observed to be misaligned by the RM effect is expected,
as the RM effect can detect much smaller degrees of misalignment than
the coarser technique described here.  Assume for the moment that the
smallest degree of misalignment this technique can identify is $|i_p -
i_s|_{min} \approx 50^{\circ}$ (as suggested by the minimum difference
from Table~\ref{tbl-2}).  Of the 26 systems with measurements of the
RM effect, six have $|\lambda| \gtrsim 50^{\circ}$.  At the same time,
the slightly higher incidence of misaligned systems in the sample of
RM measurements might be related to the fact that RM measurements are
resource intensive.  That is, systems that seem to have a higher a priori
probability of spin-orbit misalignment based on the current understanding
of spin-orbit misalignment (e.g. eccentric systems) are more likely to
be targeted for RM measurement.  The fact that this analysis includes all
known transiting exoplanet systems without the targeting biases inherent
in the current sample of RM measurements may even produce a less biased
spin-orbit misalignment distribution.

\section{Discussion}

I identify ten host stars of transiting exoplanet systems in which
either the host star is more slowly rotating relative to the simple
model than any of the stars in the control sample of SPOCS stars or in
which the star is viewed pole-on and the system is spin-orbit misaligned.
This degeneracy can be broken for at least Kepler-5 and Kepler-7, as high
precision Kepler photometry should reveal their rotation periods.  If the
periods are indeed short, then the small observed $v\sin{i}$ indicates
spin-orbit misalignment along the line of sight.  Alternatively, if the
observed rotation periods are long, then Kepler-5 and Kepler-7 are systems
in which the host star has somehow lost angular momentum much more quickly
than expected by models of Sun-like stellar spin-down.  High precision
photometry from other sources might enable similar measurements for the
other systems identified as anomalously slow projected rotators.

All ten of the candidate misaligned systems have host stars more massive
than the Sun.  To determine the significance of this observation,
I performed a Monte Carlo simulation in which I randomly selected
with replacement ten stars from the 75 listed in Table~\ref{tbl-1}.
I repeated this process 10$^6$ times.  Less than 0.01\% of the Monte
Carlo trials produced a sample in which all ten stars had $1.2~M_{\odot}
\lesssim M_{\ast} \lesssim 1.5~M_{\odot}$.  Therefore, the probability
that all ten candidate misaligned systems would be identified around
massive stars by chance is less than one in ten thousand.  Similarly,
the median mass of the planets in the candidate misaligned systems is $2
\pm 0.6$ Jupiter-masses, while the median mass of the whole sample of 75
systems is $1 \pm 0.1$ Jupiter-masses.  Interestingly, there may be a hint
of the same apparent overrepresentation of massive planets in spin-orbit
misaligned systems from Rossiter-McLaughlin measurements \citep{tor10}.

Eccentricity also seems to be related to misalignment probability.
Indeed, of the nine transiting systems with FGK stellar hosts and
eccentricity $e > 0.1$ (HD 80606, HD 17156, HAT-P-2, WASP-8, XO-3,
HAT-P-11, WASP-17, CoRoT-9, and HAT-P-14), there is now evidence from
either measurements of the RM effect or this analysis that six of those
systems are significantly misaligned (HD 80606, HD 17156, WASP-8,
XO-3, WASP-17, and HAT-P-14).  The other three systems are unlikely
to be observed to be spin-orbit misaligned for a number of reasons:
HAT-P-2 is likely an intrinsically fast-rotator and therefore unlikely
to be identified as misaligned along the line of sight by the method
described in this paper even if it does have significant spin-orbit
misalignment along the line of sight, HAT-P-11 may have too small a
radius for measurements of the RM effect, and CoRoT-9 has the largest
pericenter distance by far of the known transiting planets and almost
certainly has had a different evolutionary history than the other
known transiting planets.  To determine the significance of the apparent
correlation between eccentricity and spin-orbit misalignment, I performed
a Monte Carlo simulation to determine the expected number of systems
with eccentricity $e > 0.1$ from a sample of 16 transiting systems --
the total number of significantly misaligned systems from RM measurements
and this analysis -- expected if spin-orbit misalignment is unrelated to
eccentricity.  I randomly selected with replacement the eccentricities
of 16 planets from the 75 systems listed in Table~\ref{tbl-1}, and I
repeated this process 10$^6$ times.  I find that the mean number of
systems with $e > 0.1$ in a sample of 16 systems under the assumption
that eccentricity is unrelated to spin-orbit misalignment is $2 \pm 2$.
The observation of six systems with $e > 0.1$ from a random sample
of 16 systems occurs in less than one in 150 trials, indicating that
eccentricity and spin-orbit misalignment are related.

In an attempt to derive some constraint on the underlying spin-orbit
misalignment distribution in the 75 systems in Table~\ref{tbl-1},
I performed two completeness calculations for two different $i_s$
distributions resulting from two different $\psi$ distributions: (1)
corresponding to planet-planet scattering \citep[e.g.][]{cha08}, a case
in which the inclination of the host star is completely independent of
the alignment of the orbit of its planet $i_{s} \sim \arccos(1-U)$, where
$U \sim Unif\left(0,1\right)$; and (2) corresponding to Kozai cycles and
tidal friction \citep[e.g.][]{fab07}, a case in which the distribution
of $i_s$ is derived from the $\psi$ distribution as given in Figure
10 of \citet{fab07} and the $\lambda$ distribution as given in the RM
compilation in Table 5 of Triaud et al. (2010, submitted).  Assuming a
threshold detectable misalignment that corresponds to the smallest
degree of misalignment identified ($|i_p - i_s| \gtrsim 50^{\circ}$) in
75 systems, I find that the expected number of detections in case (1) is
$17 \pm 4$ and in case (2) it is $27 \pm 4$.  Even though each individual
system in this analysis is securely identified as an anomalously slow
projected rotator and therefore possibly spin-orbit misaligned, both
completeness calculations are strongly dependent on the smallest degree
of misalignment this technique can identify.  That quantity is uncertain,
and as such more data is necessary before any conclusive statements about
distribution of spin-orbit alignments is made.  The full sample of Kepler
transiting planet detections and host star properties available at the
end of its four year mission might resolve this issue.

Collectively, this analysis indicates that predictions of spin-orbit
misalignment for a population of close-in planets entirely produced by
planet-planet scattering or Kozai cycles with tidal friction overpredict
the number of misaligned systems.  As a result, there seems to be two
populations of close-in planets: a population that is spin-orbit aligned
and a population that is apparently spin-orbit misaligned.  The processes
that lead to misalignment or to the survival of misaligned systems seem to
operates most efficiently in systems with massive host stars and planets.
Indeed, the striking appearance of candidate misaligned systems at
$M_{\ast} \gtrsim 1.2~M_{\odot}$ in Figure~\ref{fig04} indicates that the
processes that lead to spin-orbit misalignment and survival are threshold
processes, similar to the rapid increase in exoplanet incidence with
host stellar metallicity \citep[e.g.][]{san04,fis05}.  In this case,
the transition to frequent apparent spin-orbit misalignment occurs at
the same stellar mass at which Sun-like stars with near solar metallicity
develop radiative envelopes.  This dramatic change in stellar structure
will likely strongly effect the poorly-understood tidal processes that
play a major role in the formation, evolution, and long-term survival
of spin-orbit misaligned systems.

\section{Conclusion}

There are ten transiting exoplanet systems from a larger sample of 75
systems in which the projected rotation velocity of the host star assuming
spin-orbit alignment is apparently smaller relative to a simple empirical
model of stellar spin-down than any of 866 stars in a representative
control sample of stars in the same mass range from the Spectroscopic
Properties of Cool Stars (SPOCS) catalog: HAT-P-7, HAT-P-14, HAT-P-16,
HD 17156, Kepler-5, Kepler-7, TrES-4, WASP-1, WASP-12, and WASP-14.
Unless the host stars of these transiting exoplanets are slower
rotators relative to the model prediction than any of the 866 stars
in the representative control sample, the anomalously low $v\sin{i}$
values are best explained by a pole-on view of the star and therefore
spin-orbit misalignment.  I only identify significant possible spin-orbit
misalignment in systems with host stars more massive than the Sun,
as in all ten systems the host stars have stellar masses $M_{\ast}$
in the range $1.2~M_{\odot} \lesssim M_{\ast} \lesssim 1.5~M_{\odot}$.
This observation is significant, as there is less than a one in ten
thousand chance of its occurrence by chance.  There is evidence that
systems with massive and eccentric planets are also those most likely to
be significantly spin-orbit misaligned.  Given current measurements of the
Rossiter-McLaughlin effect and frequency of misalignment along the line
of sight determined in this analysis, models of close-in planet formation
invoking planet-planet scattering or Kozai cycles and tidal friction alone
overpredict the number of misaligned systems.  As a result, there seems to
be two populations of close-in planets: those that are aligned and those
that are apparently misaligned.  The process that leads to misalignment
or survival of misaligned systems likely operates more efficiently in
systems with both massive stars and planets.  Interestingly, apparently
misaligned systems appear to occur preferentially in systems in which the
envelope of the host star is radiative, and this transition in stellar
structure is likely related through tidal processes to the formation,
evolution, and long-term survival of spin-orbit misaligned systems.

\acknowledgments I thank Greg Laughlin and Connie Rockosi for useful
comments and conversation.  This research has made use of NASA's
Astrophysics Data System Bibliographic Services.  This material is
based upon work supported under a National Science Foundation Graduate
Research Fellowship.

\clearpage

\clearpage

\clearpage

\clearpage

\clearpage

\clearpage
\LongTables
\begin{deluxetable}{lccccccccccc}
\tablecaption{Transiting Exoplanet System Properties\label{tbl-1}}
\tablewidth{0pt}
\tablehead{\colhead{System} & \colhead{$M_{\ast}$} & \colhead{$R_{\ast}$} &
\colhead{$\tau_L$} & \colhead{$\tau_U$} & \colhead{$v\sin{i}_{obs}$} &
\colhead{$\sigma_{obs}$} & \colhead{$i_p$} &
\colhead{$\overline{v\sin{i}}_{sim}$} & \colhead{$\sigma_{sim}$} &
\colhead{$\Theta$} & \colhead{Reference}\\
\colhead{} & \colhead{[$M_{\odot}$]} & \colhead{[$R_{\odot}$]} &
\colhead{[Gyr]} & \colhead{[Gyr]} & \colhead{[km s$^{-1}$]} &
\colhead{[km s$^{-1}$]} & \colhead{[deg]} & \colhead{[km s$^{-1}$]} &
\colhead{[km s$^{-1}$]} & \colhead{} & \colhead{}}
\startdata
CoRoT-1 & 0.95 & 1.11 &    $\cdots$ &    $\cdots$ &  5.20 &  1.00 & 85.10 &   2.17 &   0.87 &  -2.28 & 1 \\
CoRoT-2 & 0.97 & 0.90 &  0.20 &  4.00 & 11.85 &  0.50 & 87.84 &   3.10 &   1.27 &  -6.43 & 2,3 \\
CoRoT-3 & 1.37 & 1.56 &  1.60 &  2.80 & 17.00 &  1.00 & 85.90 &  14.92 &   1.21 &  -1.32 & 4 \\
CoRoT-4 & 1.16 & 1.17 &  0.70 &  2.00 &  6.40 &  1.00 & 90.00 &   6.23 &   0.93 &  -0.12 & 5,6 \\
CoRoT-5 & 1.00 & 1.19 &  5.50 &  8.30 &  1.00 &  1.00 & 85.83 &   2.05 &   0.12 &   1.04 & 7 \\
CoRoT-6 & 1.05 & 1.03 &  1.00 &  3.30 &  7.50 &  1.00 & 89.07 &   3.62 &   0.62 &  -3.29 & 8 \\
CoRoT-7 & 0.93 & 0.87 &  1.20 &  2.30 &  3.50 &    $\cdots$ & 80.10 &   2.57 &   0.24 &  -0.91 & 9 \\
CoRoT-9 & 0.99 & 0.94 &  0.15 &  8.00 &  3.50 &    $\cdots$ & 89.99 &   2.66 &   1.54 &  -0.46 & 10 \\
GJ 436 & 0.45 & 0.46 &  1.00 & 10.00 &  2.40 &    $\cdots$ & 85.80 &   1.03 &   0.35 &  -1.30 & 11,12,13 \\
GJ 1214 & 0.16 & 0.21 &  3.00 & 10.00 &  2.00 &    $\cdots$ & 88.62 &   5.65 &   0.96 &   2.63 & 14 \\
HAT-P-1 & 1.13 & 1.12 &  0.70 &  5.20 &  3.75 &  0.58 & 86.28 &   4.06 &   1.16 &   0.24 & 12,15 \\
HAT-P-2 & 1.36 & 1.64 &  2.10 &  3.10 & 20.80 &  0.30 & 86.72 &  13.60 &   0.77 &  -8.72 & 16 \\
HAT-P-3 & 0.94 & 0.82 &  0.10 &  6.90 &  0.50 &  0.50 & 87.24 &   2.19 &   1.24 &   1.26 & 17 \\
HAT-P-4 & 1.26 & 1.59 &  3.60 &  6.80 &  5.50 &  0.50 & 89.90 &   5.81 &   0.53 &   0.42 & 18 \\
HAT-P-5 & 1.16 & 1.17 &  1.20 &  4.70 &  2.60 &  1.50 & 86.75 &   4.31 &   0.82 &   1.00 & 12,19 \\
HAT-P-6 & 1.29 & 1.46 &  1.70 &  2.80 &  8.70 &  1.00 & 85.51 &   9.09 &   0.66 &   0.33 & 12,20 \\
HAT-P-7 & 1.47 & 1.84 &  1.20 &  3.20 &  3.80 &  0.50 & 85.70 &  32.66 &   4.61 &   6.23 & 21 \\
HAT-P-8 & 1.28 & 1.58 &  2.40 &  4.40 & 11.50 &  0.50 & 87.50 &   7.77 &   0.67 &  -4.45 & 22 \\
HAT-P-9 & 1.28 & 1.32 &  0.20 &  3.40 & 11.90 &  1.00 & 86.50 &   9.95 &   3.87 &  -0.49 & 23 \\
HAT-P-11 & 0.81 & 0.75 &  2.40 & 12.40 &  1.50 &  1.50 & 88.50 &   1.00 &   0.24 &  -0.33 & 24 \\
HAT-P-12 & 0.73 & 0.70 &  0.50 &  4.50 &  0.50 &  0.40 & 89.00 &   1.62 &   0.52 &   1.72 & 25 \\
HAT-P-13 & 1.22 & 1.56 &  4.30 &  7.50 &  2.90 &  1.00 & 83.40 &   4.54 &   0.36 &   1.54 & 26 \\
HAT-P-14 & 1.39 & 1.47 &  0.90 &  1.70 &  8.40 &  0.50 & 83.50 &  20.57 &   1.91 &   6.16 & 27 \\
HAT-P-16 & 1.22 & 1.24 &  1.20 &  2.80 &  3.50 &  0.50 & 86.60 &   6.32 &   0.78 &   3.03 & 28 \\
HD 17156 & 1.26 & 1.45 &  2.30 &  3.70 &  2.60 &  0.50 & 86.20 &   6.86 &   0.48 &   6.15 & 29,30 \\
HD 80606 & 0.97 & 0.98 &    $\cdots$ &    $\cdots$ &  1.80 &  1.00 & 89.29 &   2.08 &   0.85 &   0.21 & 31,32,33 \\
HD 149026 & 1.29 & 1.50 &  1.00 &  2.80 &  6.00 &  0.50 & 85.40 &  10.32 &   1.51 &   2.71 & 12,34,35 \\
HD 189733 & 0.81 & 0.76 &  2.40 & 12.00 &  3.32 &  0.05 & 85.76 &   1.03 &   0.23 &  -9.75 & 12,36,37 \\
HD 209458 & 1.12 & 1.16 &  2.40 &  3.90 &  4.70 &  0.16 & 86.55 &   3.66 &   0.26 &  -3.45 & 12,38 \\
Kepler-4 & 1.22 & 1.49 &  3.00 &  6.00 &  2.20 &  1.00 & 89.76 &   5.04 &   0.49 &   2.55 & 39 \\
Kepler-5 & 1.37 & 1.79 &  2.40 &  3.60 &  4.80 &  1.00 & 86.30 &  14.63 &   0.87 &   7.41 & 40 \\
Kepler-6 & 1.21 & 1.39 &  2.80 &  4.80 &  3.00 &  1.00 & 86.80 &   4.95 &   0.38 &   1.82 & 41 \\
Kepler-7 & 1.35 & 1.84 &  2.50 &  4.50 &  4.20 &  0.50 & 86.50 &  12.51 &   1.05 &   7.13 & 42 \\
Kepler-8 & 1.21 & 1.49 &  2.34 &  5.34 & 10.50 &  0.70 & 84.07 &   5.30 &   0.64 &  -5.48 & 43 \\
OGLE-TR-10 & 1.14 & 1.17 &  0.10 &  7.20 &  7.00 &  1.00 & 83.87 &   4.52 &   2.70 &  -0.86 & 12,44,45 \\
OGLE-TR-56 & 1.23 & 1.36 &  1.90 &  4.20 &  3.20 &  1.00 & 79.80 &   5.72 &   0.65 &   2.12 & 12,44,45 \\
OGLE-TR-111 & 0.85 & 0.83 &  2.20 & 14.00 &  5.00 &    $\cdots$ & 88.10 &   1.11 &   0.29 &  -3.73 & 12,44,45 \\
OGLE-TR-113 & 0.78 & 0.77 & 10.80 & 14.00 &  5.00 &    $\cdots$ & 89.40 &   0.73 &   0.03 &  -4.27 & 12,44,46 \\
OGLE-TR-132 & 1.31 & 1.32 &  0.10 &  2.70 &  5.00 &    $\cdots$ & 83.30 &  14.01 &   6.69 &   1.33 & 12,44,45 \\
OGLE2-TR-L9 & 1.52 & 1.53 &  0.00 &  0.66 & 39.33 &  0.38 & 79.80 & 128.20 & 205.72 &   0.43 & 47 \\
TrES-1  & 0.88 & 0.81 &  0.90 &  7.10 &  1.08 &  0.30 & 88.40 &   1.64 &   0.47 &   0.99 & 12,48,49 \\
TrES-2  & 0.98 & 1.00 &  2.90 &  7.70 &  2.00 &  1.00 & 83.62 &   1.92 &   0.27 &  -0.08 & 12,50,51 \\
TrES-3  & 0.93 & 0.83 &  0.10 &  3.70 &  1.50 &  1.00 & 81.85 &   2.88 &   1.53 &   0.76 & 52 \\
TrES-4  & 1.40 & 1.81 &  2.50 &  4.40 &  8.50 &  0.50 & 82.86 &  16.49 &   1.34 &   5.60 & 51,52 \\
WASP-1 & 1.30 & 1.52 &  2.40 &  3.60 &  5.00 &    $\cdots$ & 88.25 &   8.58 &   0.50 &   3.20 & 12,45,53 \\
WASP-2 & 0.89 & 0.84 &  0.00 & 14.00 &  5.00 &    $\cdots$ & 84.80 &   1.61 &   1.72 &  -1.71 & 12,51,53 \\
WASP-3 & 1.24 & 1.31 &  0.70 &  3.50 & 13.40 &  1.50 & 85.06 &   7.37 &   1.70 &  -2.66 & 54,55 \\
WASP-4 & 0.85 & 0.87 &  2.00 &  9.00 &  2.00 &  1.00 & 89.35 &   1.37 &   0.29 &  -0.61 & 56 \\
WASP-5 & 0.96 & 1.03 &  1.10 &  9.80 &  3.50 &  1.00 & 85.80 &   2.02 &   0.63 &  -1.25 & 56,57 \\
WASP-6 & 0.88 & 0.87 &  4.00 & 18.00 &  1.40 &  1.00 & 88.47 &   1.02 &   0.22 &  -0.37 & 58 \\
WASP-7 & 1.28 & 1.24 &    $\cdots$ &    $\cdots$ & 17.00 &  2.00 & 89.60 &   6.98 &   2.13 &  -3.43 & 59 \\
WASP-8 & 1.03 & 0.95 &  3.00 &  5.00 &  2.00 &  0.60 & 88.52 &   2.29 &   0.17 &   0.46 & 60 \\
WASP-10 & 0.70 & 0.78 &  0.60 &  1.00 &  6.00 &    $\cdots$ & 86.90 &   2.82 &   0.20 &  -3.12 & 61 \\
WASP-11/HAT-P-10 & 0.83 & 0.79 &  4.10 & 11.70 &  0.50 &  0.20 & 88.60 &   0.99 &   0.15 &   1.96 & 62 \\
WASP-12 & 1.35 & 1.57 &  1.00 &  3.00 &  2.20 &  1.50 & 83.10 &  14.26 &   2.19 &   4.55 & 63 \\
WASP-13 & 1.03 & 1.34 &  3.60 & 14.00 &  4.90 &    $\cdots$ & 86.90 &   2.25 &   0.44 &  -2.43 & 64 \\
WASP-14 & 1.21 & 1.31 &  0.50 &  1.00 &  2.80 &  0.57 & 84.32 &  10.45 &   1.05 &   6.39 & 65,66 \\
WASP-15 & 1.18 & 1.48 &  2.60 &  6.70 &  4.00 &  2.00 & 85.50 &   4.40 &   0.57 &   0.19 & 67 \\
WASP-16 & 1.02 & 0.95 &  0.10 &  8.10 &  3.00 &  1.00 & 85.22 &   2.71 &   1.63 &  -0.15 & 68 \\
WASP-17 & 1.20 & 1.38 &  0.40 &  3.90 &  9.00 &  1.50 & 87.80 &   6.95 &   2.25 &  -0.76 & 69 \\
WASP-18 & 1.28 & 1.23 &  0.00 &  2.00 & 11.00 &  1.50 & 86.00 &  15.25 &  14.39 &   0.29 & 70,71 \\
WASP-19 & 0.96 & 0.94 &  1.00 & 14.50 &  4.00 &  2.00 & 80.50 &   1.58 &   0.59 &  -1.16 & 72 \\
WASP-21 & 1.01 & 1.06 &  7.00 & 17.00 &  1.50 &  0.60 & 88.75 &   1.43 &   0.18 &  -0.11 & 73 \\
WASP-22 & 1.10 & 1.13 &  1.00 &    $\cdots$ &  3.50 &  0.60 & 89.20 &   3.21 &   0.95 &  -0.26 & 74 \\
WASP-24 & 1.13 & 1.15 &  0.00 &  3.70 &  6.96 &    $\cdots$ & 85.71 &   7.46 &  11.68 &   0.04 & 75 \\
WASP-25 & 1.00 & 0.95 &  0.40 &  4.60 &  3.00 &  1.00 & 87.70 &   3.06 &   1.06 &   0.04 & 76 \\
WASP-26 & 1.12 & 1.34 &  4.00 &  8.00 &  2.40 &  1.30 & 82.50 &   3.06 &   0.31 &   0.49 & 77 \\
WASP-28 & 1.08 & 1.05 &  3.00 &  8.00 &  2.20 &  1.60 & 89.10 &   2.39 &   0.33 &   0.11 & 78 \\
WASP-29 & 0.82 & 0.85 & 10.00 &    $\cdots$ &  1.50 &  0.60 & 87.96 &   0.80 &   0.06 &  -1.15 & 79 \\
WASP-33 & 1.50 & 1.44 &  0.00 &  0.50 & 90.00 & 10.00 & 87.67 & 131.62 & 200.45 &   0.21 & 80 \\
XO-1 & 1.03 & 0.93 &  0.10 &  4.10 &  1.11 &  0.67 & 89.06 &   3.76 &   1.96 &   1.28 & 12,45,81 \\
XO-2 & 0.97 & 0.98 &  3.90 &  8.70 &  1.30 &  0.30 & 88.90 &   1.69 &   0.20 &   1.08 & 82,83 \\
XO-3 & 1.21 & 1.38 &  2.00 &  3.40 & 18.54 &  0.17 & 84.20 &   5.79 &   0.44 & -27.04 & 84,85 \\
XO-4 & 1.32 & 1.56 &  1.50 &  2.70 &  8.80 &  0.50 & 88.70 &  11.76 &   1.02 &   2.61 & 86 \\
XO-5 & 1.00 & 1.11 &  7.70 &  9.30 &  0.70 &  0.50 & 86.80 &   1.71 &   0.05 &   2.01 & 87,88 \\
\enddata
\tablecomments{(1) \citet{bar08}; (2) \citet{alo08}; (3) \citet{bou08};
(4) \citet{del08}; (5) \citet{aig08}; (6) \citet{mou08};
(7) \citet{rau09}; (8) \citet{fri10}; (9) \citet{leg09};
(10) \citet{dee10}; (11) \citet{san04}; (12) \citet{tor08};
(13) \citet{bea08}; (14) \citet{cha09}; (15) \citet{joh08};
(16) \citet{pal10}; (17) \citet{tor07}; (18) \citet{kov07};
(19) \citet{bak07b}; (20) \citet{noy08}; (21) \citet{pal08};
(22) \citet{lat09}; (23) \citet{shp09}; (24) \citet{bak10};
(25) \citet{har09}; (26) \citet{bak09b}; (27) \citet{tor10};
(28) \citet{buc10}; (29) \citet{fis07}; (30) \citet{win09a};
(31) \citet{fis05}; (32) \cite{pon09b}; (33) \citet{fos09};
(34) \citet{sat05}; (35) \citet{nut09}; (36) \citet{win07};
(37) \citet{tri09}; (38) \citet{win05}; (39) \citet{bor10};
(40) \citet{koc10}; (41) \citet{dun10}; (42) \citet{lat10};
(43) \citet{jen10}; (44) \citet{mel06}; (45) \citet{sout08};
(46) \citet{gil06}; (46) \citet{sne09}; (48) \citet{lau05};
(49) \citet{nar07}; (47) \citet{soz07}; (51) \citet{dae09};
(52) \citet{soz09}; (53) \citet{cam07}; (54) \citet{pol08};
(55) \citet{gib08}; (56) \citet{gil09a}; (57) \citet{sout09a};
(58) \citet{gil09b}; (59) \citet{hel09a}; (60) Queloz et al. (2010, submitted);
(61) \citet{chr09}; (62) \citet{bak09a}; (63) \citet{heb09};
(64) \citet{ski09}; (65) \citet{jos09}; (66) \citet{joh09};
(67) \citet{wes09}; (68) \citet{lis09}; (69) \citet{and10};
(70) \citet{hel09b}; (71) \citet{sout09b}; (72) \citet{heb10};
(73) Bouchy et al. (2010, submitted); (74) Maxted et al. (2010, submitted); (75) Street et al. (2010, submitted);
(76) Enoch et al. (2010, submitted); (77) Smalley et al. (2010, submitted); (78) West et al. (2010, submitted);
(79) Hellier et al. (2010, submitted); (80) Cameron et al. (2010, submitted); (81) \citet{mcc06};
(82) \citet{bur07}; (83) \citet{fer09}; (84) \citet{kru08};
(85) \citet{win08}; (86) \citet{mcc08}; (87) \citet{bur08};
(88) \citet{pal09}}
\end{deluxetable}

\clearpage
\begin{deluxetable}{lcccccccc}
\tablecaption{Misalignment Properties\label{tbl-2}}
\tablewidth{0pt}
\tablehead{\colhead{System} & \colhead{$i_p$} & \colhead{$i_{s,1}$} &
\colhead{$i_{s,2}$} & \colhead{$i_{s,3}$} & \colhead{$i_{s,4}$} &
\colhead{$\lambda$} & \colhead{Reference}\\
\colhead{} & \colhead{[deg]} & \colhead{[deg]} & \colhead{[deg]} &
\colhead{[deg]} & \colhead{[deg]} & \colhead{[deg]} & \colhead{}}
\startdata
HAT-P-7 b & 85.70 & 6.7 & 170 & 190 & 350 & -177.5 & \citet{win09b} \\
HAT-P-14 b & 83.50 & 24 & 160 & 200 & 340 & $\cdots$ & \\
HAT-P-16 b & 86.60 & 34 & 150 & 210 & 330 & $\cdots$ & \\
HD 17156 b & 86.20 & 22 & 160 & 200 & 340 & 10 & \citet{nar09} \\
Kepler-5 b & 86.30 & 19 & 160 & 200 & 340 & $\cdots$ & \\
Kepler-7 b & 86.50 & 20 & 160 & 200 & 340 & $\cdots$ & \\
TrES-4  & 82.86 & 31 & 150 & 210 & 330 & -6.3 & \citet{nar10} \\
WASP-1 b & 88.25 & 36 & 140 & 220 & 320 & $\cdots$ & \\
WASP-12 b & 83.10 & 8.8 & 170 & 190 & 350 & $\cdots$ & \\
WASP-14 b & 84.32 & 15 & 160 & 200 & 340 & -33.1 & \citet{joh09} \\
\enddata
\tablecomments{Column 2 is the inclination of the transiting planet's
orbit to the line of sight $i_p$.  Columns 3 through 6 are the four
possible values of $i_s$ given the value of $\sin{i}$ inferred from
the difference in $v\sin{i}_{obs}$ and $\overline{v\sin{i}_{sim}}$.
Column 7 is the measured value of $\lambda$ from the Rossiter-McLaughlin
measurement taken from the indicated reference.}
\end{deluxetable}
\end{document}